\documentclass[aps,pre,twocolumn,showpacs]{revtex4}

\usepackage{amsmath,amsfonts,float}
\usepackage{graphicx}
\usepackage{epstopdf}

\begin{document}

\title{Dynamics of a Monolayer of Microspheres on an Elastic Substrate}

\author{S. P. Wallen$^{1}$, A. A. Maznev$^2$, and N. Boechler$^{1}$}

\affiliation{ 
$^1$Department of Mechanical Engineering, University of Washington, Seattle, WA, 98195 USA \\
$^2$Department of Chemistry, Massachusetts Institute of Technology, Cambridge, MA 02139 USA
}


\begin{abstract}
We present a model for wave propagation in a monolayer of spheres on an elastic substrate. The model, which considers sagittally polarized waves, includes: horizontal, vertical, and rotational degrees of freedom; normal and shear coupling between the spheres and substrate, as well as between adjacent spheres; and the effects of wave propagation in the elastic substrate. For a monolayer of interacting spheres, we find three contact resonances, whose frequencies are given by simple closed-form expressions. For a monolayer of isolated spheres, only two resonances are present. The contact resonances couple to surface acoustic waves in the substrate, leading to mode hybridization and ``avoided crossing'' phenomena. We present dispersion curves for a monolayer of silica microspheres on a silica substrate, assuming adhesive, Hertzian interactions, and compare calculations using an effective medium approximation to a discrete model of a monolayer on a rigid substrate. While the effective medium model does not account for discrete lattice effects at short wavelengths, we find that it is well suited for describing the interaction between the monolayer and substrate in the long wavelength limit. We suggest that a complete picture of the dynamics of a discrete monolayer adhered to an elastic substrate can be found using a combination of the results presented for the discrete and effective medium descriptions. This model is potentially scalable for use with both micro- and macroscale systems, and offers the prospect of experimentally extracting contact stiffnesses from measurements of acoustic dispersion.
\end{abstract}

\maketitle

\section{Introduction}

Granular media are simultaneously one of the most common and complex forms of matter on Earth. This complexity stems, in part, from heterogeneous structure and highly nonlinear particulate interactions \cite{DuranBook,GranularPhysicsBook,NesterenkoBook}. Over the past thirty years, mechanical wave propagation in ordered granular media has become an active field of research as it provides a setting for the broader understanding of granular media dynamics \cite{NesterenkoBook}. Ordered granular media have also been shown to enable a wide array of novel passive wave tailoring devices that leverage the nonlinear response stemming from the Hertzian relationship between elastic particles in contact \cite{Hertz,Johnson}, in conjunction with dispersion induced by periodicity \cite{GranularCrystalReviewChapter} or local resonances \cite{LocalResonance}.

Experimental configurations used to study mechanical wave propagation in ordered granular media typically involve spherical particles confined by elastic media. This type of arrangement is particularly common in one- and two-dimensional configurations, and includes macro- to microscale particles. For example, at the macroscale, elastic rod structures, tracks, and tubes have been used to confine the particles in one-dimensional \cite{LocalResonance,Macro1D,MacroUpshift} and quasi-one-dimensional \cite{MacroQuasi1D} configurations, and elastic plates have been used in two dimensions \cite{Macro2D}. More recently, the dynamics of a two-dimensional monolayer of 1 $ \mu\mathrm{m} $ silica particles adhered to an elastic substrate was studied using a laser ultrasonic technique \cite{Boechler_PRL}.

Analytical models used to describe the dynamics of these systems typically only include the interaction between the particles (often just the normal Hertzian contact interaction) and disregard the effect of the substrate. In reality, even for the simple case of a particle monolayer on a substrate, more complex dynamics involving interactions between the particles and elastic waves in the substrate should be expected. Indeed, a recent experiment \cite{Boechler_PRL} showed that a monolayer of microspheres adhered to a substrate strongly interacts with Rayleigh surface waves in the substrate, leading to the hybridization between Rayleigh waves and a microsphere contact resonance. The results of this experiment were analyzed with a simple model involving only vertical (normal to the substrate surface) vibrations of isolated particles, following the approach adopted in earlier theoretical works on the interaction of surface oscillators with Rayleigh waves \cite{Baghai1992,Garova1999}. However, in reality, the particle motion is not confined to the vertical direction, and the Rayleigh wave has a significant horizontal component. Furthermore, the interaction between neighboring particles in contact is expected to significantly influence the dynamics.      

A notable theoretical work \cite{Kosevich1989} provided a model for the dynamics of a monolayer adhered to an elastic substrate which accounted for both normal and horizontal motion and interaction between the particles. However, this study did not take into account particle rotation. A more recent study \cite{Tournat2011} demonstrated that the rotational degree of freedom has a profound effect on the dynamics of granular monolayers. However, the analysis of monolayers on  substrates in Ref. \cite{Tournat2011} only accounted for normal contact interactions between the particles and the substrate, and the substrate was considered rigid. 

The aim of this work is to provide a theoretical model for the contact-based dynamics of a two-dimensional layer of spheres on a substrate, accounting for the elasticity of the substrate, translational and rotational motion of the spheres, and both normal and shear stiffnesses of sphere-to-sphere and sphere-substrate contacts. We focus on a system with microscale particles that interact with each other and with the substrate via van der Waals adhesion forces. Rather than postulate the contact stiffness constants, we derive them from Hertzian contact models. This imposes certain constraints on the values of the constants: for example, the ratio of the normal and shear contact stiffness between the spheres is a constant only weakly dependent on Poisson’s ratio. We consider contact-based modes having frequencies significantly below the intrinsic spheroidal vibrational modes of the spheres, such that they can be described as spring-mass oscillators. Furthermore, we focus on dynamics involving particle and substrate displacements in the sagittal plane, as would be detectable in a laser-based experiment, such as that of Ref. \cite{Boechler_PRL}.  

We start with the case of a rigid substrate, where we find three eigenmodes involving vertical, horizontal and rotational motion of the spheres. In the long-wavelength limit these modes yield three contact resonances, for which simple analytical expressions are obtained. One of the resonances only involves motion of the spheres normal to the substrate surface, whereas the other two involve mixed horizontal-rotational motion. We then present our effective medium model, which describes the interaction between the spheres and the substrate. The results show that the contact resonances interact with Rayleigh surface waves, which leads to mode hybridization and avoided crossings.  We discuss the behavior for cases involving both isolated (non-touching) and interacting spheres, and demonstrate the important role of rotations in both cases.  We also examine the validity of the effective medium approximation, by comparing the calculations using discrete and effective medium models. Finally, we discuss the implications of our findings for past and future studies on granular monolayer systems.

\hspace{\textwidth}
\section{Model}
	\begin{figure}[]
		\centering
		\includegraphics[width=0.9\columnwidth]{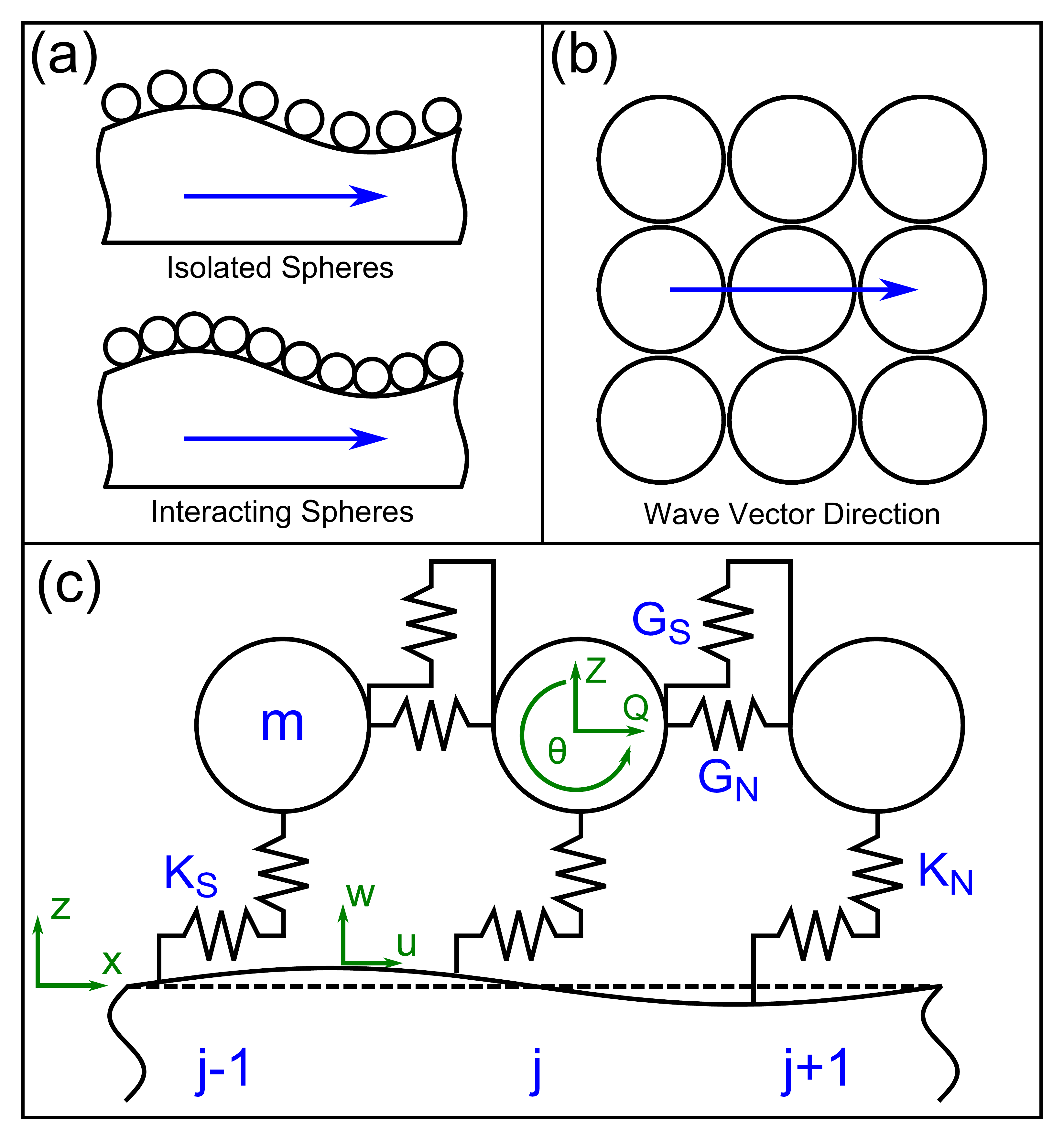}
		\caption{(a) Side-view schematic of an amplified wave profile for isolated and interacting spheres. (b) Top-down view of the square-packed monolayer, with the arrow indicating the direction of wave propagation. (c) Schematic for the model of a monolayer of spheres coupled to an elastic halfspace.}
		\label{Model_Schem}
	\end{figure}
	We consider a monolayer of elastic spheres on a substrate, which can be either close-packed and in contact, or isolated, as shown in Fig. \ref{Model_Schem}(a). In either case, the spheres are assumed to form a square lattice, with the wave propagation direction aligned with the lattice vector, as shown in Fig. \ref{Model_Schem}(b). We model the layer as an infinite lattice of rigid spheres with diameter \(D = 2R\) and mass \(m\), coupled to a semi-infinite, isotropic elastic substrate by normal and shear stiffnesses \(K_N\) and \(K_S\), and to nearest-neighbor spheres by stiffnesses \(G_N\) and \(G_S\), as schematically shown in Fig. \ref{Model_Schem}(c). The subscript \(N\) corresponds to forces acting normal to the surface of the sphere, and \(S\) to forces acting transverse to the surface of the sphere. The shear springs generate an associated torque about the sphere center, while the normal springs do not. The absolute horizontal, vertical, and angular displacements of sphere \(j\) from its equilibrium state are given by \(Q_j\), \(Z_j\), and \(\theta_j\), respectively, and the displacements of the substrate are denoted by \(u(x,z)\), corresponding to displacement in the \(x\)-direction, and \(w(x,z)\), corresponding to displacements in the \(z\)-direction. 
	\subsection{Contact Stiffness} \label{Contact}
		We derive the stiffnesses $K_N$, $K_S$, $G_N$, and $G_S$ using Derjaguin-Muller-Toporov (DMT) \cite{DMT1983, Israelachvili} and Mindlin contact models \cite{Mindlin}. The DMT theory is typically applicable in weakly-adhesive systems with small, stiff particles \cite{Bhushan}, and assumes that the deformation profile is Hertzian. The Mindlin model describes the shear stiffness of particles in contact, assuming an applied normal force \cite{Mindlin}. At the microscale, adhesive contact models have been explored experimentally in the quasi-static regime using atomic force microscopy and nanoindentation approaches \cite{Fuchs}.
		
		For contact between two spheres (or a sphere and a halfspace) having elastic moduli \(E_1\) and \(E_2\), and Poisson's ratios \(\nu_1\) and \(\nu_2\), the Hertzian restoring elastic force \(F_N\) corresponding to displacement \(\delta_N\) of the particle center in the direction normal to the contact surface is given by	
		%
		\begin{align}
		F_N &= \frac{4}{3} E^* R_{c}^{1/2} \delta_N^{3/2},\label{contactForceN}
		\end{align}
		%
		\noindent
		where \(R_{c}\) is the effective radius of contact (equal to \(R\) for sphere-sphere contacts and \(R/2\) for sphere-halfspace contacts), and \(E^* = [(1-\nu_1^2)/E_1 + (1-\nu_2^2)/E_2]^{-1}\) is the effective modulus. Considering the DMT adhesive force \(F_{DMT} = 2 \pi w R_c\) acting normal to the contact surface \cite{DMT1983, Israelachvili} (where \(w\) is the work of adhesion between two surfaces), the net normal force is given by
		%
		\begin{equation}
		F_{N,net} = F_N - F_{DMT}.
		\label{contactForceNnet}
		\end{equation} 
		To describe the shear contact, we utilize the Mindlin model \cite{Mindlin}, which assumes small relative displacements and no slip at the contact surface. For two elastic bodies with shear moduli \(G_1\) and \(G_2\), the restoring elastic force \(F_S\) to displacement \(\delta_S\) of the particle center in the direction transverse to the contact normal is given by \cite{Mindlin}
		\begin{align}
		F_S &= 8 G^* R_{c}^{1/2} \delta_S \delta_N^{1/2},\label{contactForceS}
		\end{align} 
		\noindent
		where \(G^* = [(2-\nu_1)/G_1 + (2-\nu_2)/G_2]^{-1}\) is the effective shear modulus. Here, the factor of \(\delta_N^{1/2}\) arises from the Hertzian relation between the contact radius and \(F_N\).
		
		By substituting the relative displacements \(\delta_N = Z - w_0\) and \(\delta_S = Q - u_0 + R\theta\) into Eqs. \ref{contactForceNnet} and \ref{contactForceS}, and linearizing about the equilibrium configuration of \(\delta_{N,0} = [3 F_{DMT}/(4 E^* R_{c}^{1/2})]^{2/3}\) and  \(\delta_{S,0} = 0\), we derive linearized normal and shear contact stiffnesses
		\begin{align}
		\begin{split}
		K_N &=\left(6 E^{*^2} R_c F_{DMT}\right)^{1/3}\\
		K_S &= 8 \left(\frac{3}{4} \frac{G^{*^3}}{E^*} R_c F_{DMT}\right)^{1/3},\label{springConst}
		\end{split}
		\end{align}
		\noindent
		with $ G_N $ and $ G_S $ given by equations of the same form, but with $ E^* $, $ G^* $, $ R_c $, and $F_{DMT}$ adjusted for sphere-sphere contacts. In the special case where the spheres and substrate are composed of the same material, the relative magnitudes of the stiffness constants are determined exclusively by Poisson's ratio $ \nu $ of the material, 
		\begin{align}
		\begin{split}
		G_N &= 2^{-2/3} K_N\\
		K_S &= \nu^* K_N\\
		G_S &= 2^{-2/3} \nu^* K_N,\label{springConstReduced}
		\end{split}
		\end{align}
		\noindent
		where \(\nu^* = 2(1-\nu)/(2-\nu)\). \hfill\\*
%
%
		\hspace{\textwidth}
	\subsection{Equations of Motion of the Spheres}
		Assuming small displacements (i.e. $Q_j $, $ Z_j $, and $ R \theta_j $ are much less than $ D $), the \(j^{th}\) sphere obeys the equations of motion
		\begin{align}
		\begin{split}
			m \ddot{Q}_j = & - K_S(Q_j - u_{0,j} + R \theta_j)\\ 
						   & + G_N(Q_{j+1} - 2 Q_j + Q_{j-1})\\
			m \ddot{Z}_j = & - K_N(Z_j - w_{0,j})\\
						   & + G_S[Z_{j+1} - 2 Z_j + Z_{j-1} - R(\theta_{j+1} -\theta_{j-1})]\\
			I \ddot{\theta}_j = & - K_S R(Q_j - u_{0,j} + R \theta_j)\\
							    & - G_S R [R(\theta_{j+1} + 2 \theta_j + \theta_{j-1})\\
							    & - (Z_{j+1} - Z_{j-1})], \label{EOM_Disc}
		\end{split}
		\end{align}
		\noindent where \(u_{0,j}\) and $ w_{0,j} $ are horizontal and vertical displacements of the substrate surface at the point of contact, respectively.
		\hspace{\textwidth}
	\subsection{Effective Medium Approximation}
		Considering wavelengths much longer than the sphere diameter, we treat the monolayer as an effective continuous medium. By substituting the center difference formulas \([(\cdotp)_{j+1} - (\cdotp)_{j-1}]/(2D) \approx \partial (\cdotp) / \partial x\) and \([(\cdotp)_{j+1} - 2(\cdotp)_j + (\cdotp)_{j-1}] / (D^2) \approx \partial^2 (\cdotp) / \partial x^2\) into Eq. (\ref{EOM_Disc}), we arrive at the equations of motion of the monolayer in effective medium form:
		\begin{align}
		\begin{split}
		m \frac{\partial^2 Q}{\partial t^2} = & - K_S(Q - u_0 + R \theta)\\ 
		& + 4 G_N R^2 \frac{\partial^2 Q}{\partial x^2}\\
		m \frac{\partial^2 Z}{\partial t^2} = & - K_N(Z - w_0)\\
		& + 4 G_S R^2 (\frac{\partial^2 Z}{\partial x^2} - \frac{\partial \theta}{\partial x})\\
		I \frac{\partial^2 \theta}{\partial t^2} = & - K_S R(Q - u_0 + R \theta)\\
		& - 4 G_S R^2 (R^2 \frac{\partial^2 \theta}{\partial x^2} + \theta + \frac{\partial Z}{\partial x} ). \label{EOM_Cont}
		\end{split}
		\end{align}
		The coupling between the monolayer and substrate is described by the following boundary conditions at the surface \(z = 0\), which describe the average force acting on the surface due to the motion of the spheres:
		\begin{align}
		\begin{split}
		\sigma_{zx} &= \frac{K_S}{A}(Q - u_0 + R \theta)\\
		\sigma_{zz} &= \frac{K_N}{A}(Z - w_0),\label{BC}
		\end{split}
		\end{align} 
		\noindent where \(\sigma_{zx}\) and \(\sigma_{zz}\) are components of the elastic stress tensor \cite{Ewing} and \(A\) = \(D^2\) is the area of a primitive unit cell in our square-packed monolayer. The combination of Eq. (\ref{EOM_Cont}) and the linear elastic wave equations describing waves in the substrate \cite{Ewing}, coupled by the boundary conditions of Eq. (\ref{BC}), comprises the complete effective medium model.
%
\hspace{\textwidth}
\section{Dispersion Relations} \label{DispRel}
	\subsection{Rigid Substrate}
%
		\subsubsection{Discrete Model}
			We first consider the discrete model, which accurately captures the structural periodicity of the monolayer. We assume spatially-discrete traveling wave solutions of the form \(\hat{Q} e^{i(\omega t - kDj)}\) (with similar terms for the other displacements) and set the displacements of the substrate surface \(u_{0,j}\) and \(w_{0,j}\) to zero (rigid substrate). Here, \(\hat{(\cdotp)}\) is the amplitude of a plane wave in the displacement variable \((\cdotp)\), \(\omega\) is the angular frequency, and \(k\) is the wave number. After algebraic manipulation, Eq. (\ref{EOM_Disc}) is reduced to a homogeneous system of three linear algebraic equations in the amplitudes \(\hat{(\cdotp)}\). This system possesses non-trivial solutions only for pairs of $ k $ and $ \omega $ that cause the determinant of the system to vanish. Enforcing this condition, we arrive at the following dispersion relation, where the three rows of the determinant correspond to the three equations of Eq. (\ref{EOM_Disc}): 
			\begin{widetext}
				\begin{equation}
				\left|\begin{array}{ccc}
				\frac{c_N^2}{2 R^2}(1-\cos(kD)) + \phi_S \omega_S^2 & 0 &  R\omega_S^2\\
				0 & \frac{c_S^2}{2 R^2}(1-\cos(kD)) + \phi_N \omega_N^2 & -\frac{c_S^2}{2 R} i\sin(kD)\\
				 R\omega_S^2 & \frac{c_S^2}{2 R} i\sin(kD) & \frac{I}{m} [\frac{c_\theta^2}{2 R^2}(1-\cos(kD)) + \phi_\theta \omega_\theta^2]
				\end{array}\right| = 0,
				\label{DRdet_disc}
				\end{equation}
				\hspace{\textwidth}
			\end{widetext}
			\noindent
			where \(\phi_N = 1 - \omega^2 / \omega_N^2\), \(\phi_S = 1 - \omega^2 / \omega_S^2\), \(\phi_\theta = 1 - \omega^2 / \omega_\theta^2\), \(c_N=\sqrt{G_N / m}(2R)\) and \(c_S=\sqrt{G_S / m } (2R)\) are the longitudinal and transverse long-wavelength sound speeds of the discrete monolayer, \(c_\theta^2 = -mR^2c_S^2 / I\), \(\omega_N^2 =K_N / m\), \(\omega_S^2 =K_S / m\), and \(\omega_\theta^2 = (K_S + 4G_S)R^2 / I\).
			\hspace{\textwidth}
		\subsubsection{Effective Medium}
			In the effective medium model, which approximates the discrete system at long wavelengths, we substitute spatially-continuous traveling wave solutions of the form \(\hat{Q} e^{i(\omega t - kx)}\) (with similar terms for the other displacements) into \(Q\), \(Z\), and \(\theta\) in Eq. (\ref{EOM_Cont}) with \(u_{0} = w_{0} = 0\). Following the same procedure as in the discrete model, we arrive at the dispersion relation for the effective medium model:

			\begin{widetext}
				\begin{equation}
					\left|\begin{array}{ccc}
					c_N^2 k^2 + \phi_S \omega_S^2 & 0 & R \omega_S^2\\
					0 & c_S^2 k^2 + \phi_N \omega_N^2 & -ik c_S^2\\
					R\omega_S^2 & ikc_S^2 & \frac{I}{m}(c_\theta^2 k^2 + \phi_\theta \omega_\theta^2)
					\end{array}\right| = 0.
					\label{DRdet_cont_RB}
				\end{equation}
				\hspace{\textwidth}
			\end{widetext}
			
			It is particularly instructive to examine the behavior of the effective medium model in the long wavelength limit $k \rightarrow 0$. In this limit, the displacements vary slowly in space, and the spatial derivative terms of Eq. (\ref{EOM_Cont}) may be neglected. For the case of a rigid base, Eq. (\ref{EOM_Cont}) then reduces to the form
			
			\begin{align}
			\begin{split}
			m \frac{\partial^2 Q}{\partial t^2} = & - K_S(Q + R \theta)\\ 
			m \frac{\partial^2 Z}{\partial t^2} = & - K_N Z\\
			I \frac{\partial^2 \theta}{\partial t^2} = & - K_S R(Q + R \theta) - 4 G_S R^2 \theta. \label{EOM_Cont_noSD}
			\end{split}
			\end{align}
			
			\noindent
			The equation for $Z$ decouples from the other two equations and yields a vertical vibrational mode. The two other equations remain coupled, yielding two modes containing both horizontal and rotational motion. Using the moment of inertia of a solid sphere \(I = (2/5) m R^2\), we find three resonance frequencies
			\begin{align}
			\begin{split}
			\omega_N & = \left[\frac{K_N}{m}\right]^{1/2}\\
			\omega_{RH} & = \left[\left(\frac{K_S}{4m}\right) \left(20 \gamma + 7 + \sqrt{400 \gamma^2 + 120 \gamma + 49}\right)\right]^{1/2}\\
			\omega_{HR} & = \left[\left(\frac{K_S}{4m}\right) \left(20 \gamma + 7 - \sqrt{400 \gamma^2 + 120 \gamma + 49}\right)\right]^{1/2}. \label{Res_longwave}
			\end{split}
			\end{align}
			\noindent
			where $ \gamma = G_S / K_S $. Here, \(\omega_N\) corresponds to a mode with exclusively vertical motion, described by Eq. (\ref{EOM_Cont_noSD}). The other two modes $ \omega_{RH} $ and $ \omega_{HR} $ exhibit both rotational and horizontal (but not vertical) motion, with relative amplitudes determined by $ \gamma $. The higher of the two horizontal-rotational modes is predominantly rotational and the lower is predominantly horizontal, hence we have used the notations $\omega_{RH}$ and $\omega_{HR}$, where the first letter in the subscript denotes the dominant motion. If the spheres and substrate are made of the same material, then, by using Eq. (\ref{springConstReduced}), we can relate the horizontal-rotational frequencies of Eq. (\ref{Res_longwave}) to the vertical resonance frequency, with the expressions $\omega_{RH} = 3.018 \nu^{*^{1/2}} \omega_N$ and $\omega_{HR} = 0.832 \nu^{*^{1/2}} \omega_N$.
			
\begin{figure}
				\centering
				\includegraphics[width=\columnwidth]{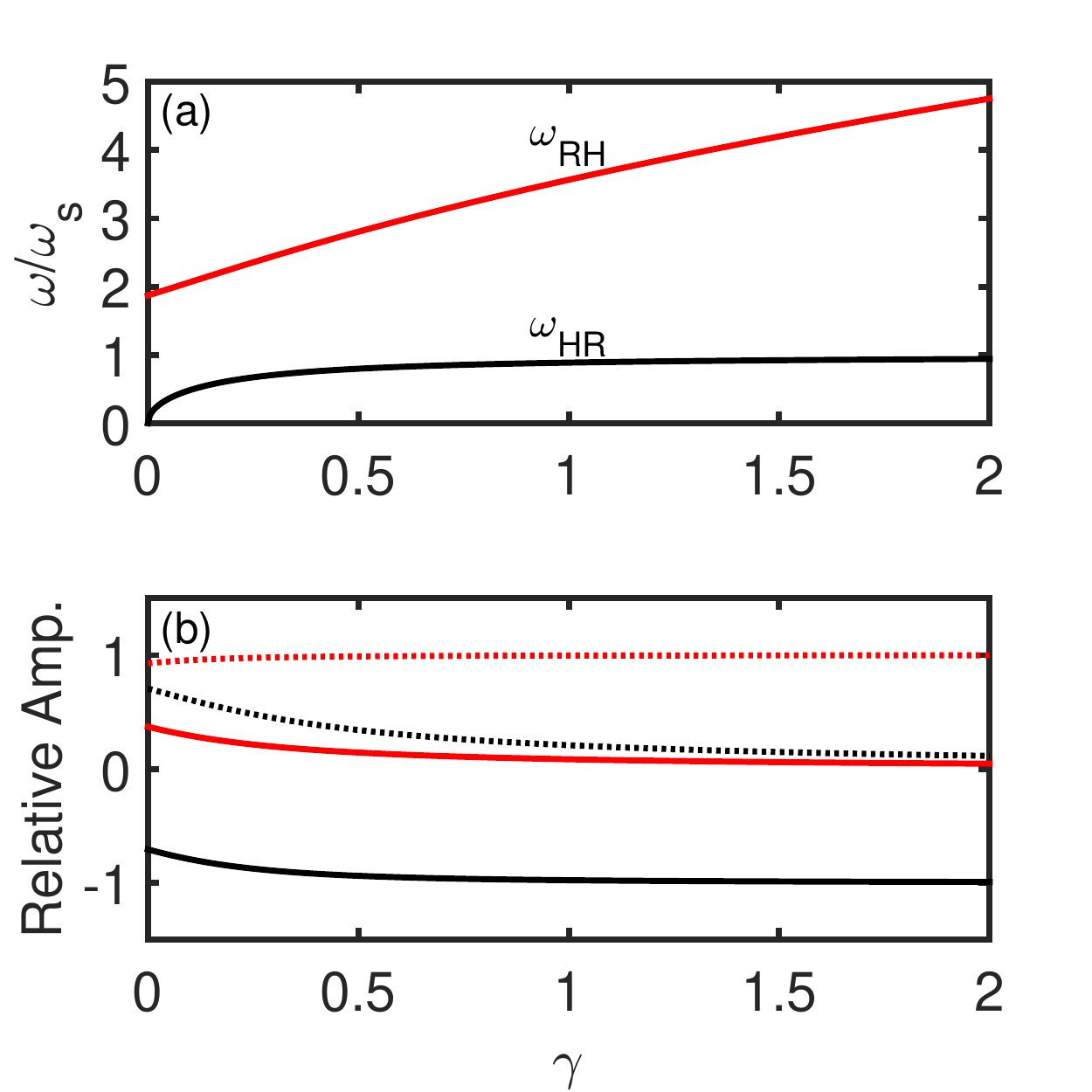}
				\caption{(a) Resonance frequencies $ \omega_{RH} $ (red line) and $ \omega_{HR} $ (black line)  as functions of the stiffness ratio $ \gamma $. (b) Displacement amplitudes of the resonant modes with frequencies $\omega_{RH} $ (red lines) and $ \omega_{HR} $ (black lines), as functions of the stiffness ratio $ \gamma $. Solid and dotted lines correspond to $ Q $ and $ R\theta $, respectively. For each resonance, the amplitudes are normalized such that the sum of squares is unity. The positive sign of $ R \theta $ corresponds to counterclockwise rotation.}  
				\label{gammaPlot}
			\end{figure}

			In the limiting case of isolated spheres (described by $ \gamma = 0 $), $ \omega_{RH} $ and $ \omega_{HR} $ of Eq. (\ref{Res_longwave}) become $ \omega_{RH,Iso} = \sqrt{7/2} \hspace{1 pt}\omega_S $ and $ \omega_{HR,Iso} = 0 $, respectively. For identical materials, $ \omega_{RH,Iso} = \sqrt{7\nu^*/2} \hspace{1 pt}\omega_N $. The dependence of $ \omega_{RH} $ and $ \omega_{HR} $ on $ \gamma $ is shown in Fig. \ref{gammaPlot} (a), where it can be seen that $ \omega_{RH} $ originates at $ \omega_{RH,Iso} $ for $ \gamma = 0 $ and grows unbounded, while $ \omega_{HR} $ originates at $ \omega_{HR,Iso} = 0 $, and approaches $ \omega_S $ asymptotically. In Fig. \ref{gammaPlot} (b), we plot the horizontal and rotational displacement amplitudes as functions of $ \gamma $ for these two modes. Different signs of the rotational amplitude indicate that the $\omega_{RH}$ and $\omega_{HR}$ modes have different motion patterns. In the former, a positive displacement is accompanied by a counterclockwise rotation, while in the latter, it is accompanied by a clockwise rotation.  
			
			We note that the zero-frequency mode, $\omega_{HR,Iso}$, corresponds to the rolling motion of an isolated sphere. With the inclusion of a bending rigidity, the sphere would not be allowed to freely roll, and instead would undergo rocking motion of a finite frequency. While non-zero bending rigidity is expected to exist in real systems (for instance, a microsphere adhered to a substrate does not freely roll), the frequency of resulting rocking vibrations is predicted to be orders of magnitude lower than the other contact resonances discussed here \cite{Tielens}. Bending rigidity would thus act as a small perturbation to the predictions of our model, and we do not include it in our analysis. 
			
			To illustrate the importance of particle rotations in the model, we note that in the limiting case of \(I \rightarrow \infty\), when there is no rotation, Eq. (\ref{EOM_Cont_noSD}) yields two resonances: a vertical resonance with frequency $\omega_N$, and a horizontal resonance having frequency \(\omega_S\). For isolated spheres, the effect of rotations can be thought of as a reduction of the ``effective mass" of the sphere to $ (2/7)m $, which increases the horizontal resonance frequency. For interacting spheres, on the other hand, rotations drastically change the dynamics, yielding two horizontal-rotational modes whose frequencies depend on the relative strengths of the sphere-to-sphere and sphere-substrate interactions.  
			
		\subsection{Elastic Substrate}
			As in the case of the effective medium approximation for a rigid substrate, we assume traveling wave solutions of the form \(\hat{Q} e^{i(\omega t - kx)}\) (with similar terms for the other displacements) into \(Q\), \(Z\), and \(\theta\) in Eq. (\ref{EOM_Cont}). Likewise, we express the variables \(u_0\), \(w_0\), \(\sigma_{zx}\) \(\sigma_{zz}\) in terms of surface wave solutions for the elastic potentials \cite{Ewing} \(\phi(x,z,t) = \hat{\phi} e^{k \alpha z + i(\omega t - kx)}\) and \(\psi(x,z,t) = \hat{\psi} e^{k \beta z + i(\omega t - kx)}\), and then substitute these expressions into Eq. (\ref{EOM_Cont}) and Eq. (\ref{BC}). Here, \(\hat{(\cdotp)}\) is the amplitude of a plane wave in the displacement or potential variable \((\cdotp)\), \(\alpha = \sqrt{1 - \omega^2/(c_L^2 k^2)}\), \(\beta = \sqrt{1 - \omega^2/(c_T^2 k^2)}\), and $ c_L $ and \(c_T\) are the longitudinal and transverse sound speeds of the substrate, respectively. After algebraic manipulation, Eq. (\ref{EOM_Cont}) and Eq. (\ref{BC}) are reduced to a homogeneous system of five linear algebraic equations in the five plane wave amplitudes \(\hat{(\cdotp)}\), with coefficients depending on \(k\) and \(\omega\). We reach the dispersion relation by seeking nontrivial solutions of this system, which exist only for pairs of \(k\) and \(\omega\) that cause the determinant of the following coefficient matrix to vanish:
			\begin{widetext}
				\begin{equation}
				\left|\begin{array}{ccccc}
				ik\omega_S^2 & k \beta \omega_S^2 & c_N^2 k^2 + \phi_S \omega_S^2 & 0 & R \omega_S^2\\
				-k \alpha \omega_N^2 & ik \omega_N^2 & 0 & c_S^2 k^2 + \phi_N \omega_N^2 & -ik c_S^2\\
				ikR\omega_S^2 & kR\beta\omega_S^2  & R\omega_S^2 & ikc_S^2 & \frac{I}{m}(c_\theta^2 k^2 + \phi_\theta \omega_\theta^2)\\
				1+\beta^2 & -2i\beta & 0 & \frac{m}{\rho A c_T^2 k^2}(c_S^2 k^2 + \phi_N \omega_N^2 - \omega_N^2) & \frac{-m}{\rho A c_T^2 k^2}ikc_S^2\\
				-2i\alpha & -(1+\beta^2) & \frac{m}{\rho A c_T^2 k^2}(c_N^2 k^2 + \phi_S \omega_S^2 - \omega_S^2) & 0 & 0
				\end{array}\right| = 0,
				\label{DRdet}
				\end{equation}
				\hspace{\textwidth}
			\end{widetext}
			
			\noindent 
			where $ \rho $ is the density of the substrate, and $ A = D^2 $ is the area of a primitive unit cell in our square-packed monolayer. We note that the coupling between the spheres and the substrate is represented by elements \((4,4)\), \((4,5)\), and \((5,3)\) of the matrix in Eq. (\ref{DRdet}). Thus, the strength of the coupling can be quantified by the ratio \(m / (\rho A)\); if this term is made to vanish (e.g. by making the mass of each sphere much less than that of the portion of the substrate below it, extending to the depth of material influenced by Rayleigh waves), then the substrate and monolayer will be effectively decoupled. We note that if rotations are disregarded (e.g. by letting $ I \rightarrow \infty $), Eq. (\ref{DRdet}) reduces to the same form as that of the adsorbed monolayer of Ref. \cite{Kosevich1989}.
			
			It is instructive to consider the long-wave limit when the spatial derivatives in Eq. (\ref{EOM_Cont}) can be disregarded. In this case, we find the simplified dispersion relation
			
			\begin{widetext}
				\begin{equation}
				\left|\begin{array}{ccccc}
				ik\omega_S^2 & k \beta \omega_S^2 & \phi_S \omega_S^2 & 0 & R \omega_S^2\\
				-k \alpha \omega_N^2 & ik \omega_N^2 & 0 & \phi_N \omega_N^2 & 0\\
				ikR\omega_S^2 & kR\beta\omega_S^2  & R\omega_S^2 & 0 & \frac{I}{m}\phi_\theta \omega_\theta^2\\
				1+\beta^2 & -2i\beta & 0 & \frac{m}{\rho A c_T^2 k^2}(\phi_N \omega_N^2 - \omega_N^2) & 0\\
				-2i\alpha & -(1+\beta^2) & \frac{m}{\rho A c_T^2 k^2}(\phi_S \omega_S^2 - \omega_S^2) & 0 & 0
				\end{array}\right| = 0.		
				\label{DRdet_reduced}
				\end{equation}
				\hspace{\textwidth}
			\end{widetext}
			\hspace{\textwidth}

			For isolated spheres, there is no approximation in Eq. (\ref{DRdet_reduced}) with respect to Eq. (\ref{DRdet}), because in this case the terms generated by the spatial derivatives in Eq. (\ref{EOM_Cont}) are identically zero. For interacting spheres, the accuracy of dispersion relations calculated with Eq. (\ref{DRdet_reduced}) will be assessed below by a comparison with results obtained with Eq. (\ref{DRdet}). We will see that Eq. (\ref{DRdet_reduced}) essentially describes the interaction of contact resonances given by Eq. (\ref{EOM_Cont_noSD}) with Rayleigh surface waves.
		
		\section{Numerical Results and Discussion}
		
		In the following calculations, we consider silica spheres of $1.08$ $\mu$m diameter on a silica substrate, and use the elastic constants (Ref. \cite{GlassProp}) \(E =\) 73 GPa, \(\nu =\) 0.17, and work of adhesion (Ref. \cite{Israelachvili}) \(w =\) 0.063 J/m\textsuperscript{2}.	
		
		\subsection{Rigid Substrate}
		
		\begin{figure}
			\centering
			\includegraphics[width=\columnwidth]{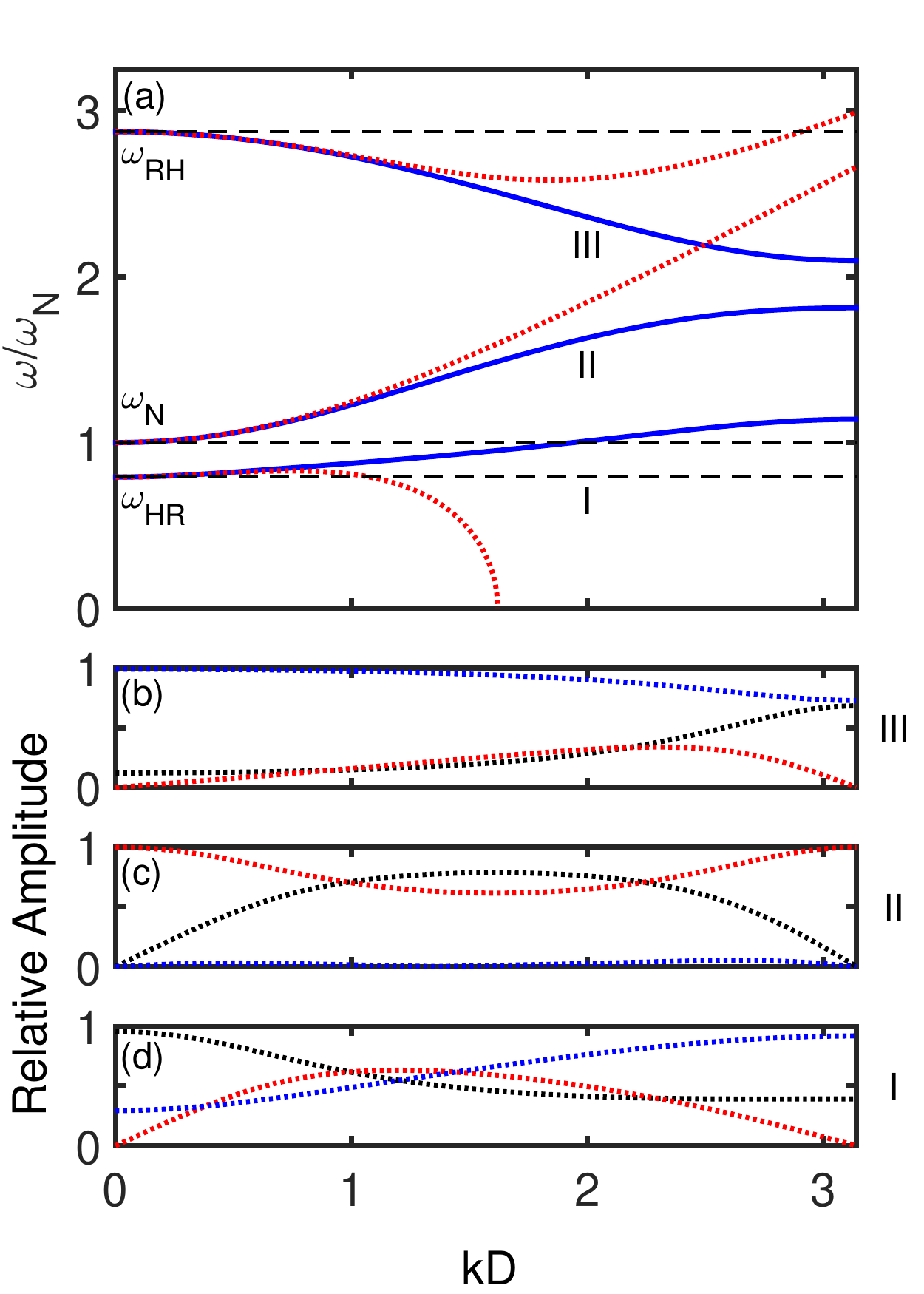}
			\caption{(a) Dispersion relation of a discrete monolayer adhered to a rigid base. Blue solid and red dotted lines denote, respectively, discrete and effective medium monolayers. Black dashed lines denote the contact resonances. (b)-(d) relative amplitudes of the displacement variables \(Q\) (black dotted lines), \(Z\) (red dotted lines), and \(R\theta\) (blue dotted lines), corresponding to the branches of the same numeral for the dispersion of the discrete monolayer adhered to the rigid base shown in (a). The amplitudes are normalized such that the sum of the squares is unity.}
			\label{DR_Layer}
		\end{figure}
		
		We plot numerical solutions of Eq. (\ref{DRdet_disc}), to obtain the dispersion curves for the discrete model of interacting spheres on a rigid base, as shown in Fig. \ref{DR_Layer}(a). In our description of a rigid substrate, we assume that no elastic waves propagate in the substrate, but allow local deformation at the points of contact for the purpose of the contact stiffness calculation; this preserves the same contact stiffnesses as in the elastic substrate analysis. We note that due to the periodicity of the system, all three branches have zero-group velocities at the edge of the first irreducible Brillouin zone \cite{BrillouinBook} of the monolayer. 
		
		By substituting the solutions shown in Fig. \ref{DR_Layer}(a) into the coefficient matrix of the corresponding algebraic system, we numerically determine the amplitudes of the sphere displacements, which we plot in Fig. \ref{DR_Layer}(b-d). By comparing the calculated displacements of with the dispersion curves, we see that each branch takes on the character of its respective contact resonance in the limit $k \rightarrow 0$. One can see that each of the three modes generally involves both vertical and horizontal, as well as rotational motion (albeit the rotational component of mode II is quite small).  The existence of the three modes with mixed displacements is a consequence of the inclusion of the rotational degree of freedom: without rotations, there would be two modes, one vertical and one horizontal.
		
		We note that, in the special case $ K_S = 0 $, the mode originating at $ \omega_{HR} $ becomes purely horizontal and decouples from the other two modes. The remaining modes (characterized by vertical translation and rotation) are generally consistent with the results of Ref. \cite{Tournat2011}, for the case of normal contact with a rigid surface and no bending rigidity. Since Ref. \cite{Tournat2011} considered hexagonal packing, the behavior is analogous at long wavelengths, but diverges at short wavelengths due to discrete lattice effects. 
		
\begin{figure}[t]
			\centering
			\includegraphics[width=\columnwidth]{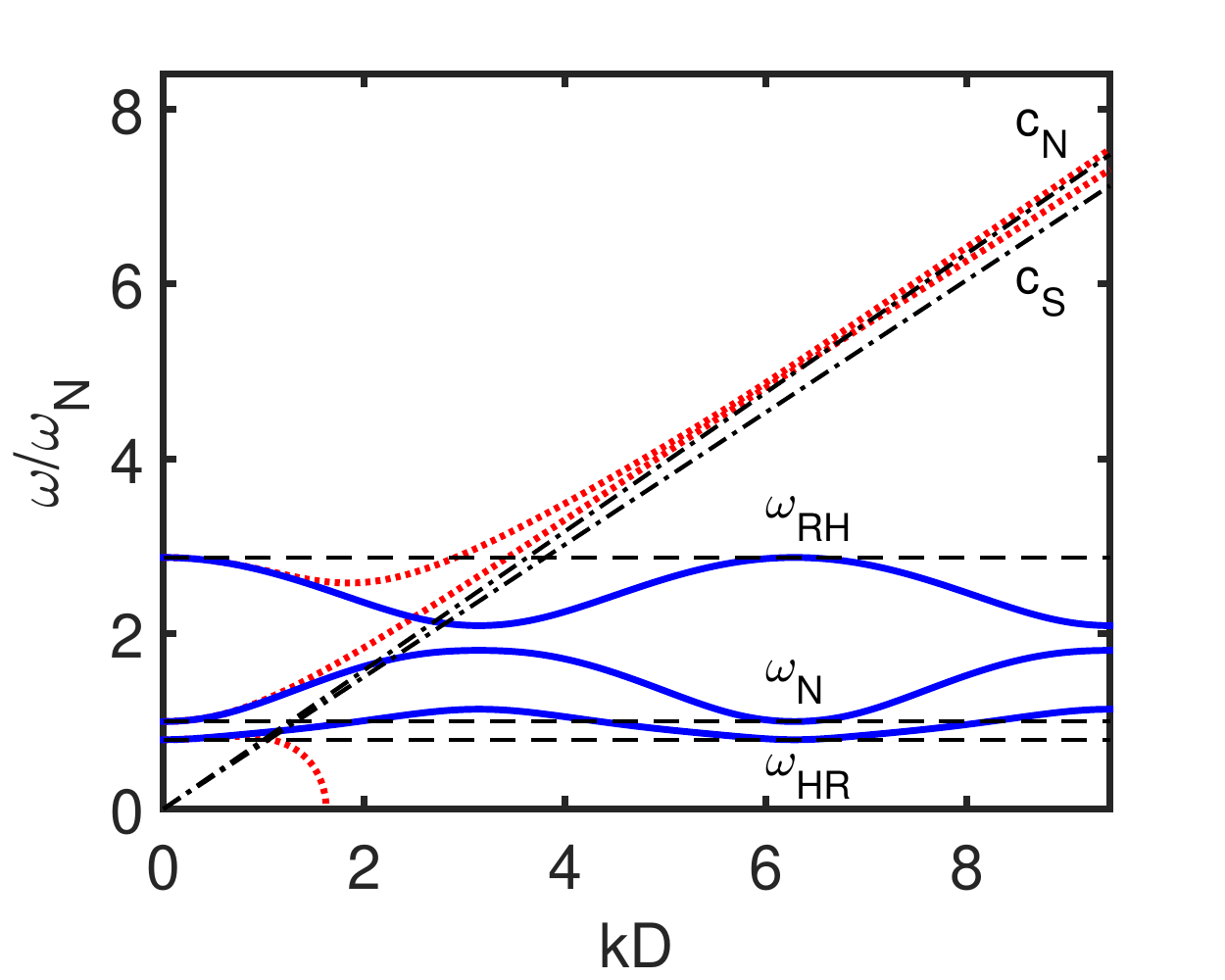}
			\caption{Dispersion relation of a monolayer adhered to a rigid base, using an extended plotting range. Blue solid, red dotted, and black dashed lines are the same as in Fig. \ref{DR_Layer}(a). Black dash-dotted lines denote the long-wavelength longitudinal and transverse sounds speeds of the monolayer.}
			\label{DR_EM_Compare}
		\end{figure}

		The dotted lines in Fig. \ref{DR_Layer}(a) show dispersion curves calculated with the effective medium model as per Eq. (\ref{DRdet_cont_RB}). The effective medium approximation yields accurate results at long wavelengths but fails at shorter wavelengths with the unphysical behavior of the first mode, whose frequency goes to zero. At even shorter wavelengths, as shown in Fig. \ref{DR_EM_Compare}, the effective medium dispersion curves of modes II and III asymptotically approach straight lines with slopes given by the longitudinal and transverse sound speeds in the monolayer. This asymptotic behavior has been described by Kosevich and Syrkin \cite{Kosevich1989}. However, as can be seen from the dispersion curves generated using the discrete model in Fig. \ref{DR_EM_Compare}, this asymptotic behavior does not occur in our system due to the spatial periodicity of the monolayer. As a result, the inclusion of the first- and second-order spatial derivative terms of Eq. (\ref{EOM_Cont}) does not deliver much additional understanding of the dynamics of our system. 
%
						
		\hspace{\textwidth}
		
		\subsection{Elastic Substrate}
		
%
		\subsubsection{Isolated Spheres} \label{iso}
		\begin{figure}[h]
			\centering
			\includegraphics[width=\columnwidth]{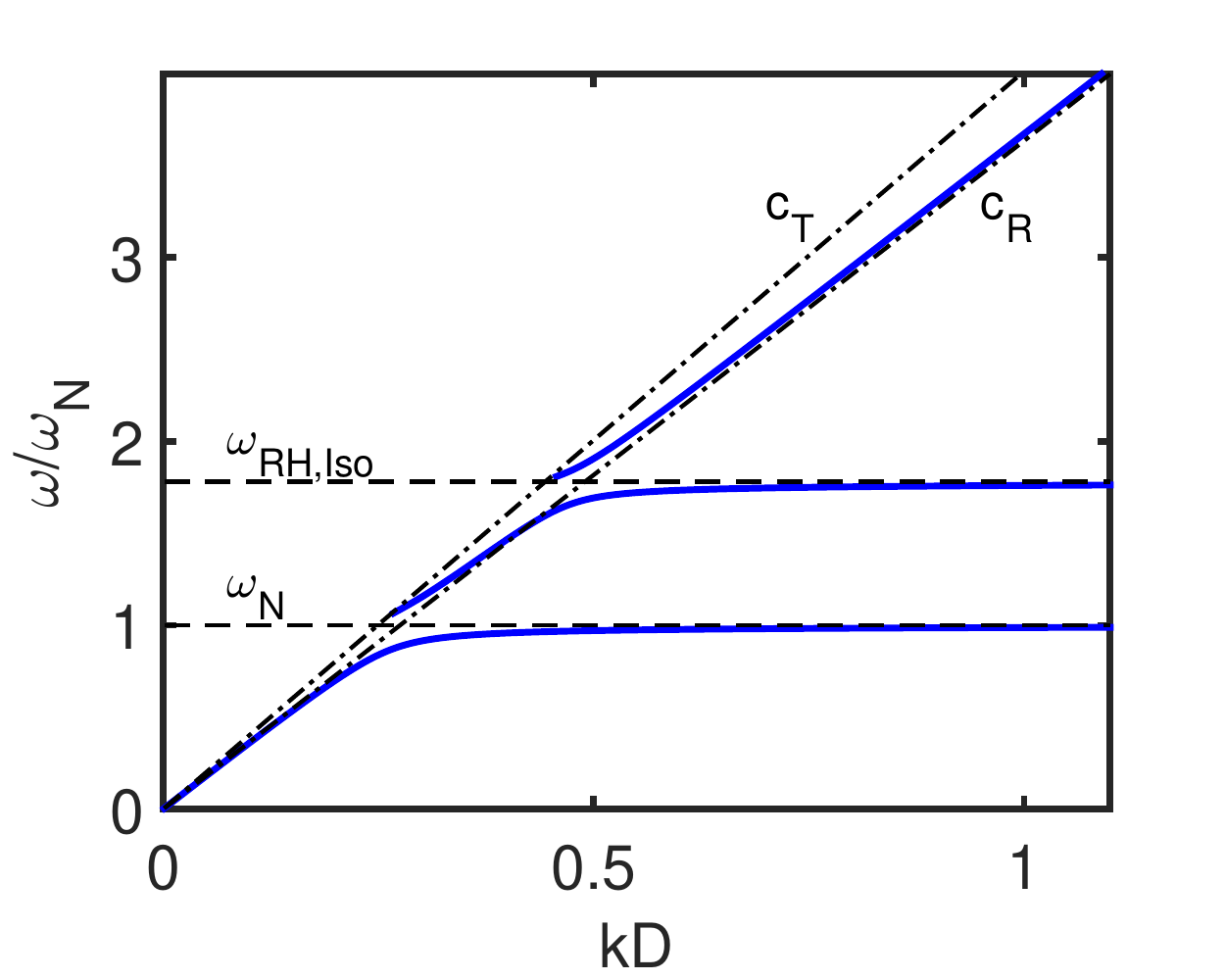}
			\caption{Dispersion relation of SAWs in an elastic half space coupled to a monolayer of isolated elastic spheres, denoted by the blue solid lines. Black dashed lines denote the contact resonances, and black dash-dotted lines denote the transverse and Rayleigh waves speeds of the substrate.}
			\label{DR_Iso}
		\end{figure}
				
		We numerically solve Eq. (\ref{DRdet_reduced}) for the isolated spheres case using \(G_S = 0\) and all other parameters derived in Sec. \ref{Contact}, and plot the resulting dispersion relation for the effective medium model, as shown in Fig. \ref{DR_Iso}. This dispersion relation exhibits classic ``avoided crossing" behavior \cite{Wigner} about the resonance frequencies \(\omega_N\) and \(\omega_{RH,Iso} = \sqrt{7/2} \hspace{2 pt}\omega_S\). In this model, surface acoustic waves (SAWs) in the substrate behave as classical Rayleigh waves at frequencies far from the contact resonances, and the dispersion curves follow the line corresponding to the substrate Rayleigh wave speed \(c_R\) \cite{Ewing}. Conversely, sphere motions dominate those of the substrate at frequencies close to the contact resonances. For phase velocities greater than \(c_T\), which correspond to the region \(\omega > c_T k\), the wave numbers that solve Eq. (\ref{DRdet}) are complex valued; these solutions are ``leaky" modes for which energy leaves the surface of the substrate, and radiates into the bulk. 
		This isolated spheres case is particularly applicable in systems where adhesion between particles is negligible, e.g. for: macroscale particles without lateral compression where the dominant static compression is due to gravity and is between the particles and substrate; or for microscale particles, if the separation distance between particles is larger than the range of adhesion forces. 
		
		\subsubsection{Interacting Spheres}
		%
		\begin{figure}
			\centering
			\includegraphics[width=\columnwidth]{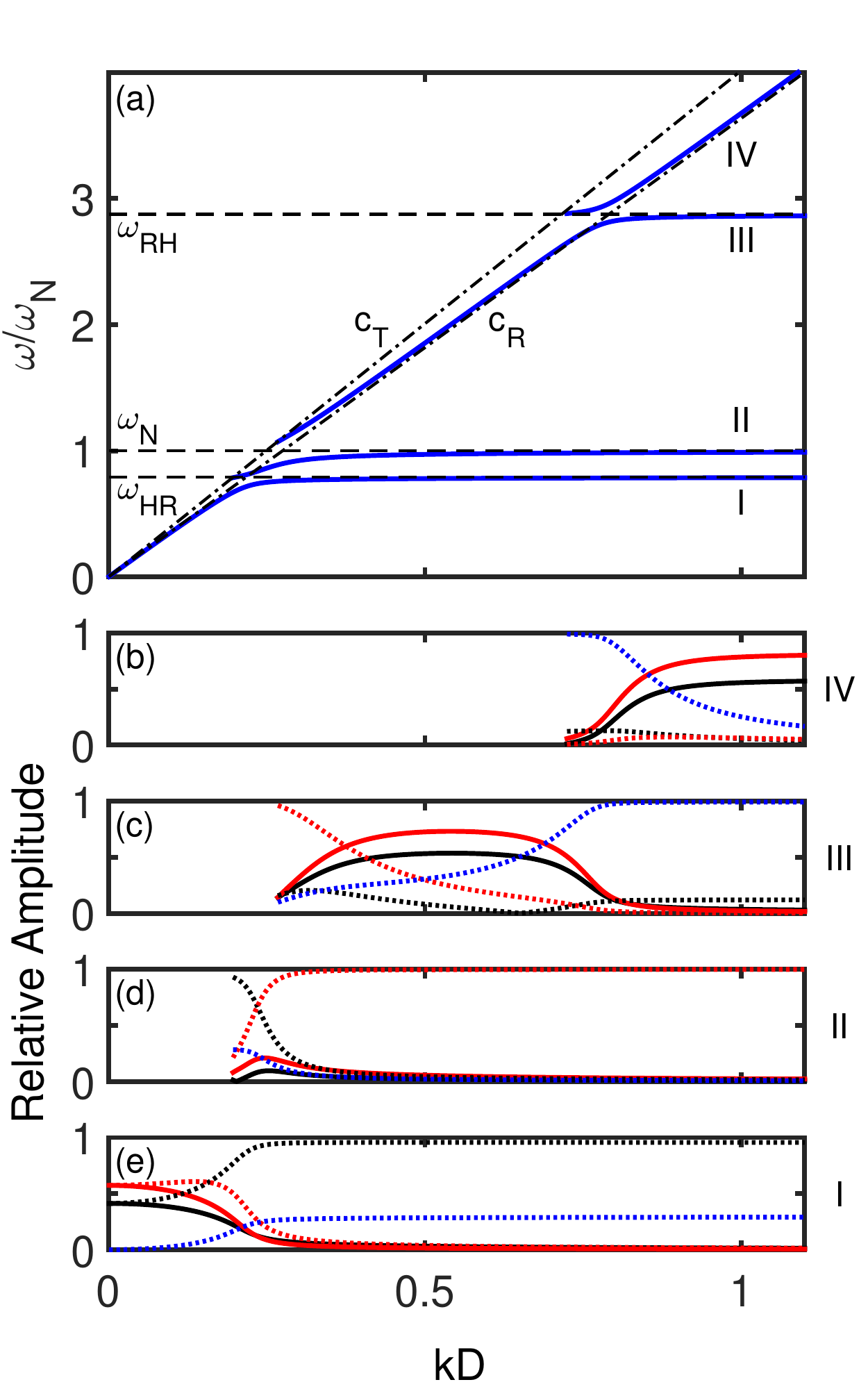}
			\caption{(a) Blue solid lines denote the dispersion relation of SAWs in an elastic halfspace coupled to a monolayer of interacting elastic spheres. Black dashed lines denote the contact resonances, and black dash-dotted lines denote wave speeds in the substrate. (b)-(e) relative amplitudes of the displacement variables \(u_0\) (black solid lines),  \(w_0\) (red solid lines),  \(Q\) (black dotted lines), \(Z\) (red dotted lines), and \(R\theta\) (blue dotted lines), corresponding to the branch denoted by the same numeral in (a). The amplitudes are normalized such that the sum of the squares is unity.}
			\label{DR_Inter}
		\end{figure}
		\noindent
		
		In Fig. \ref{DR_Inter}(a), we plot numerical solutions of Eq. (\ref{DRdet_reduced}) for the long wavelength limit of the effective medium model with interacting spheres. The amplitudes of the sphere and substrate displacements are calculated in the same manner as in Fig. \ref{DR_Layer}, and are plotted in Fig. \ref{DR_Inter}(b-e). In Fig. \ref{DR_Inter}(a), we observe features qualitatively similar to the dispersion relation for isolated spheres in Fig. \ref{DR_Iso}, with the exception of a third avoided crossing at frequency \(\omega_{HR}\). The mode shapes reveal the ways in which each of the branches are influenced by the contact resonances, as well as long and short wavelength asymptotic behavior of our system. In the long wavelength limit, the substrate motions closely resemble Rayleigh SAWs \cite{Ewing}, with a mix of vertical and horizontal motions. Since the frequencies of waves in this regime are well below the contact resonances, the effect of the spheres is negligible, and the monolayer moves in phase with the substrate surface. At short wavelengths, it can be clearly seen that the first, second, and third lowest branches exhibit motions dominated by the displacements \(Q\), \(Z\), and \(\theta\), respectively (each corresponding to a resonant mode of the monolayer), while the highest branch tends toward the Rayleigh SAW. The effects of proximity to the contact resonances are well illustrated, for example, by branch III of Fig. \ref{DR_Inter}(a), which exhibits large vertical sphere motions at its starting point near \(\omega_N\), resembles the Rayleigh SAW as it approaches and crosses the \(c_R\) line, and transitions into large rotational sphere motions after bending around the avoided crossing with \(\omega_{RH}\). 

\begin{figure}[]
			\centering
			\includegraphics[width=\columnwidth]{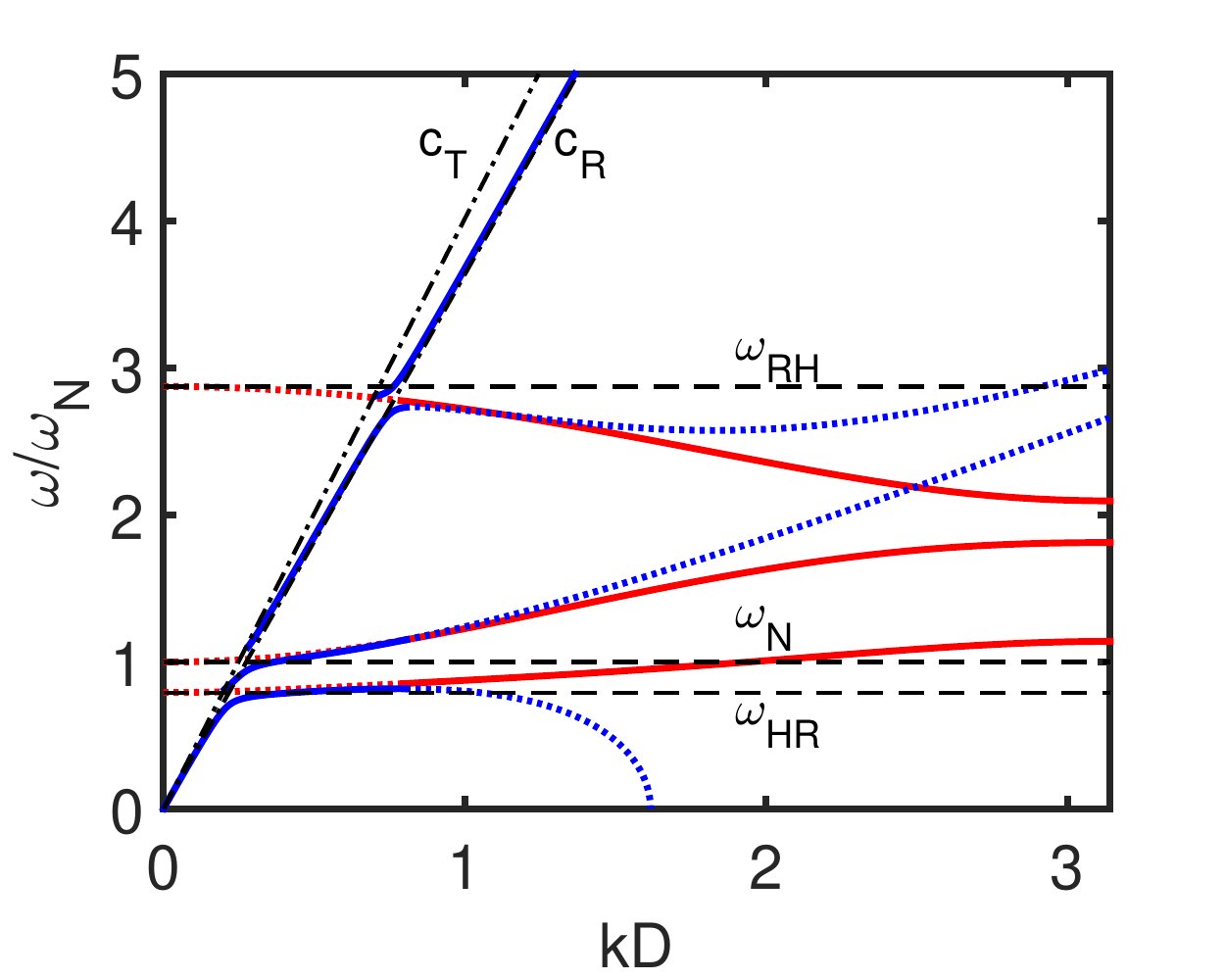}
			\caption{Blue lines denote the SAW dispersion relation with spatial derivative terms included and red lines denote the dispersion relation of a discrete monolayer adhered to a rigid base. Solid and dotted lines denote, respectively, valid and invalid ranges for the two models. Black dashed and dash-dotted lines are the same as in Fig. \ref{DR_Layer}(a).}
			\label{DR_RE_Compare}
		\end{figure}

		In order to examine the behavior of our system throughout the entire Brillouin zone, we superimpose the dispersion curves for the effective medium model of interacting spheres on an elastic base including higher order spatial derivative terms (the full Eq. (\ref{DRdet})) with the dispersion curves for the discrete monolayer on a rigid substrate (Eq. (\ref{DRdet_disc})), as shown in Fig. \ref{DR_RE_Compare}. At long wavelengths the discreteness of the monolayer is insignificant, and the dispersion is well described by the effective medium model. Furthermore, we note that at long wavelengths the dispersion curves calculated using the effective medium model including higher order terms shown in Fig. \ref{DR_RE_Compare}, only slightly deviates from the dispersion calculated using the effective medium model with the higher order terms neglected shown in Fig. \ref{DR_Inter}(a). The only noticeable effect is a downshift in frequency of the avoided crossing between the Rayleigh wave and the $\omega_{RH}$ resonance; since the latter intersects at the highest wave vector of the three contact resonances, calculations with Eq. (\ref{DRdet_reduced}) in this case are the least accurate. In Fig. \ref{DR_RE_Compare}, at short wavelengths, beyond the avoided crossings with the Rayleigh wave branch, the elasticity of the substrate has little effect on the dynamics, and the dispersion can be described using the discrete model for interacting spheres on a rigid substrate. We suggest that by ``stitching together'' the effective medium model for spheres on an elastic substrate with the discrete model for spheres on a rigid substrate, we can simultaneously capture the interaction of SAWs with the monolayer at long wavelengths and effects caused by the discreteness of the spheres at short wavelengths. Past the avoided crossings, the two sets of curves in Fig. \ref{DR_RE_Compare} stitch together smoothly, resulting in a full picture of the monolayer dynamics on the elastic substrate. 

		\hspace{\textwidth}
		
		\section{Experimental Implications}
		
		We expect the presented results to be useful for predicting complex dynamic responses and extracting effective contact stiffnesses from measurements of acoustic dispersion in a manner similar to Boechler {\it et al.} \cite{Boechler_PRL}. The findings described above invite several questions, including whether our model of a square lattice is applicable to results on hexagonally packed monolayers, and why horizontal-rotational resonances were not observed in the experiment \cite{Boechler_PRL}. 
		
		We believe that the assumption of the square lattice is not essential. For isolated spheres, Eqs. (\ref{EOM_Cont}) and (\ref{BC}) with $ G_N $ and $ G_S $ set to zero can be obtained for any arrangement of the spheres, periodic or random, with the only parameter depending on the arrangement being the surface area per sphere A. For interacting spheres, the results generally do depend on the lattice structure and the propagation direction. However, the contact resonances given by Eq. (\ref{Res_longwave}) correspond to the $ k=0 $ limit and, consequently, do not depend on the propagation direction. The relative positions of the three contact resonances may be different in the long wavelength limit between a hexagonal and square packed lattice, but their presence should still be expected in both cases. 
		
		We suggest that the reasons why horizontal-rotational resonances were not observed in Ref. \cite{Boechler_PRL} may be the following. Since the measurements were not sensitive to horizontal motion, the $ \omega_{RH} $ and $ \omega_{HR} $ resonances could only be detected when they hybridized with SAWs near avoided crossings, and since the avoided crossings with $ \omega_{RH} $ and $ \omega_{HR} $ resonances are more narrow than the one with the $ \omega_N $ resonance, they could have been missed. Furthermore, our model assumes that all spheres are either connected by identical springs or are isolated. If the contact stiffness between spheres were to vary widely (some neighboring spheres being in contact and others not, for example), then distinct resonances may be absent. In addition, the upper ($ \omega_{RH} $) resonance may have been outside the range of the measurements in Ref. \cite{Boechler_PRL}. Further experimental studies of monolayer dynamics in conjunction with exploration of ways to control sphere-to-sphere contacts should help resolve the discrepancy between the theory and experiment.
		
		While the main focus of this work has been on microgranular monolayers, our theory is equally valid for macroscale systems. In this case, the contact springs would be determined by gravity and, possibly, applied lateral static compression \cite{MacroUpshift}, rather than by adhesion forces. We note that several past experimental works on macroscale granular systems \cite{MacroUpshift} have observed systematic upshifts in frequency relative to theoretical predictions, and have suggested uncertainties in material parameters and experimental setups, as well as deviations from Hertzian contact behavior as possible causes. As per the results from our model, the presence of additional degrees of freedom and interactions between spheres and substrate may also be possible causes. In the absence of the external lateral compression, highly nonlinear ``sonic vacua'' \cite{NesterenkoBook} should also be expected. Generally, as amplitudes are increased, interesting nonlinear dynamics are expected for both micro- and macroscale monolayers due to nonlinearity of Hertzian contacts between the particles \cite{NesterenkoBook,GranularCrystalReviewChapter} and between the particles and the substrate \cite{ISOT2014}. 
				
		\section{Conclusion}
		
		We have developed a model for wave propagation in granular systems composed of a monolayer of spheres on an elastic substrate. Our model expands on those used in previous works by including the elasticity of the substrate, horizontal and rotational sphere motions, shear coupling between the spheres and substrate, and interactions between adjacent spheres. We have shown that a monolayer of interacting spheres on a rigid substrate supports three modes involving vertical, horizontal, and rotational motion. In the long-wavelength limit, these modes yield three contact resonances, one purely vertical and two of mixed horizontal-rotational character. On an elastic substrate, these resonances hybridize with the Rayleigh surface wave yielding three avoided crossings. For isolated spheres, the frequency of the lower horizontal-rotational resonance, in the absence of bending rigidity, tends to zero and only two contact resonances with two respective avoided crossings remain. 
		
		By comparing the effective medium (valid for long wavelengths) to the discrete formulation of our model, we have demonstrated that for the presented microsphere monolayer example, the effective medium model can be used to describe the interaction of the contact resonances with the Rayleigh waves in the substrate, but loses accuracy at shorter wavelengths. In that case, the substrate can be considered rigid, and the discrete model is more appropriate. This model is scalable in that it can be adapted for use with both macro- and microscale systems, and provides a means to experimentally extract contact stiffnesses from dynamic measurements. Opportunities for future studies include exploration of analogous models for granular monolayers in the nonlinear regime, as well as analysis of the transverse modes of a monolayer of spheres on an elastic substrate (the latter involves transverse horizontal displacement and rotations of the spheres, as well as shear horizontal acoustic waves in the substrate). Further experiments with macro- and microscale granular monolayers will help guide the modeling effort. 
		
		\section{Acknowledgments}
		The authors greatly appreciate discussions with Vitaly Gusev. S.W. and N.B. gratefully acknowledge support from the US National Science Foundation (grant no. CMMI-1333858) and the US Army Research Office (grant no. W911NF-15-1-0030). The contribution by A.A.M. was supported by the National Science Foundation Grant No. CHE-1111557.
		\hspace{\textwidth}

\end{document}